\documentclass[traditabstract]{aa}
\usepackage{amsmath}

\usepackage{txfonts}
\usepackage{natbib}
\bibpunct{(}{)}{;}{a}{}{,} 

\usepackage{graphicx}  
\usepackage{hyperref}
\usepackage{listings}

\usepackage{color}
\definecolor{gray}{rgb}{0.4,0.4,0.4}
\definecolor{darkblue}{rgb}{0.0,0.0,0.6}
\definecolor{cyan}{rgb}{0.0,0.6,0.6}
\definecolor{maroon}{rgb}{0.5,0,0}
\definecolor{darkgreen}{rgb}{0,0.5,0}
\lstdefinelanguage{XML}
{
  morestring=[b]",
  morestring=[s]{>}{<},
  morecomment=[s]{<?}{?>},
  stringstyle=\color{black},
  identifierstyle=\color{darkblue},
  keywordstyle=\color{cyan},
  morekeywords={xmlns,version,type}
}

\newcommand{\AM}{{A\&M}}

\begin{document}


\title{Inside a VAMDC data node}

\subtitle{Putting standards into practical software}
\titlerunning{VAMDC data node}

\author{Samuel Regandell\inst{1}
  \and Thomas Marquart\inst{1}
  \and Nikolai Piskunov\inst{1}
}

\offprints{S. Regandell, \email{samuel.regandell@physics.uu.se}} \institute{Department of Physics and Astronomy, Uppsala, Sweden}
\date{Accepted 2017-12-18}

\abstract {Access to molecular and atomic data is critical for many
forms of remote sensing analysis across different fields. Many atomic
and molecular databases are however highly specialized for their
intended application, complicating querying and combination data
between sources. The Virtual Atomic and Molecular Data Centre, VAMDC,
is an electronic infrastructure that allows each database to register
as a "node". Through services such as VAMDC's portal website, users
can then access and query all nodes in a homogenized way. Today all
major Atomic and Molecular databases are attached to VAMDC.

This article describes the software tools we developed to help data
providers create and manage a VAMDC node. It gives an overview of the
VAMDC infrastructure and of the various standards it uses. The article
then discusses the development choices made and how the standards are
implemented in practice. It concludes with a full example of implementing
a VAMDC node using a real-life case as well as future plans for the
node software.}


\maketitle



\section{Introduction}
\label{sec:introduction}

The  problem of analyzing remote sensing data is encountered in a wide variety
of scientific and technological fields, ranging from astronomy and climate
research to computer tomography and movie production. Its solution requires
detailed knowledge about absorption, scattering/reflection and emissivity of
matter.  This in turn builds on insight to the electronic and nuclear structure
of atomic and molecular species, its dependence on the environment, as well as
radiative and collisional processes leading to transitions between energy
levels. Ray tracing, spectral analysis and other types of remote sensing tools
are well-established in research and industry. By contrast, the collection of
relevant atomic and molecular data has remained challenging. This is due to the
heterogenous nature of individual data collections: Atomic and Molecular (\AM)
databases are often created to serve very specific applications (e.g.~modelling
of stellar atmospheres) or as repositories storing the output from experimental
or theoretical work by atomic physicists. Thus a particular data collection is
often specialised not only in terms of content but also in terms of units,
format and query language.

The Virtual Atomic and Molecular Data Centre (VAMDC, \url{http://vamdc.eu}) is an
electronic infrastructure that solves these problems by providing standardized
ways to both formulate queries and to retrieve data. All the data collections that are
part of VAMDC, henceforth called \textbf{data nodes}, provide their internally
heterogenous data in a commonly agreed-to, machine-redable form. VAMDC was
originally created with support from an EU FP7 grant and is now maintained and
further developed by a large international consortium. It offers atomic-
or molecular data providers the possibility to quickly bring their results to their users. At
the same time it provides the attributes of published work through including the valid references for
proper citations. All major \AM\ databases are either integrated in VAMDC or support
standards and protocols developed within this project.

\begin{figure}[h!]
  \centering
  \includegraphics[width=0.40\textwidth]{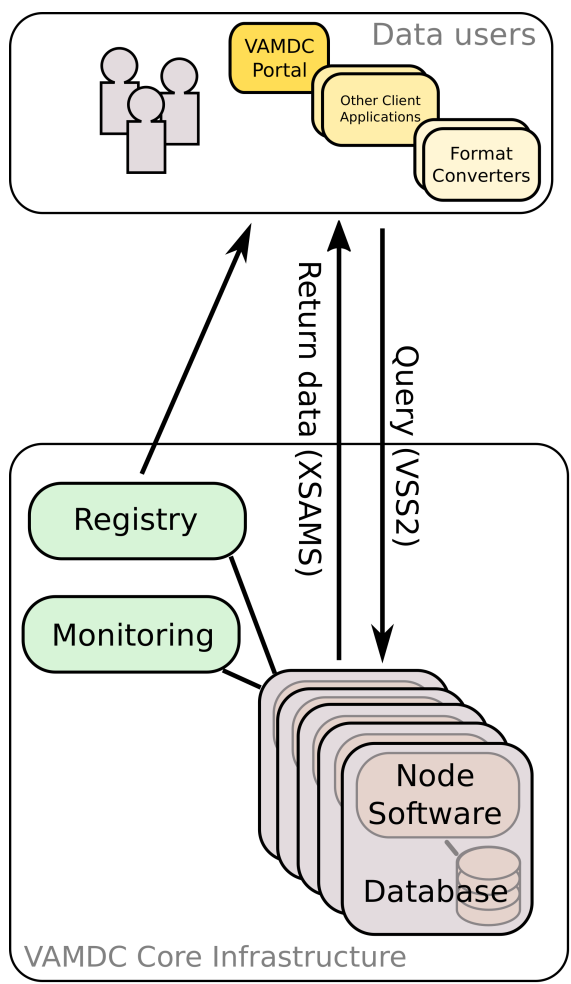}
  \caption{Graphical overview of the VAMDC infrastructure.}
  \label{fig:vamdcstructure}
\end{figure}

To make uniform access to data possible, VAMDC maintains an infrastructure of
web services. The VAMDC \textbf{Registry} lists data nodes and other services
together with their respective capabilities. This allows users and programs to
find resources and only select those containing the sort of data they need.
Most visible from the perspective of the data-user community is the VAMDC
{\textbf{portal}}\footnote{\url{http://portal.vamdc.eu/}}. The portal is a web
interface that helps with formulating queries and collecting data from the connected
data nodes. The portal also
acts as an intermediary to services that process the machine-readable results; these 
services can perform calculations and export to various file formats or human-readable
presentations. For a full overview, see \citet{vamdc2016}.

In this paper we focus on VAMDC's data provider infrastructure, consisting of
the aforementioned data nodes. A node is hosted and maintained either by the
data producer or by a group interested in collecting data from several sources.
Each node may offer a wide variety of scientific data depending on its history
and purpose. Regardless of its internal layout and content, a node can join
VAMDC by implementing the interface defined by the VAMDC standards. This 
interface must:

\begin{itemize}
    \item Understand queries formulated using the VAMDC query language.
    \item Produce data output in the standardized data format (XSAMS)
    used by VAMDC for data exchange.
    \item Implement the VAMDC application programming interface (API) to receive
    queries, to send back the results and to report additional information to clients 
    and to the registry.
\end{itemize}

The \textbf{node software} is the software package supplied by VAMDC. It aims to 
help individual data nodes set up and operate.  In the following we will first go into some
detail on the VAMDC standards.  We then describe the node software
implementation and the intricacies that arise from balancing versatility,
simplicity of use and performance at the same time.  Finally we take an
existing data node as an example to illustrate the node software in action and
the process of using the co-bundled {publishing tools} to import and
update node data content.

\section{The VAMDC Standards for Data Nodes}
\label{sec:vamdc_standards}

The VAMDC standards are publicly available online at
\mbox{\url{http://standards.vamdc.eu/}} and we refer the reader to that resource for
all details omitted here. The purpose of this section is to summarise the parts
of the standards that are particularly relevant to the node software and to
highlight a few aspects that directly affect the implementation, as described
in the subsequent sections.

\subsection{The data format: XSAMS}
\label{sec:output_language_xsams}

\begin{figure}[h!]
  \centering
  \includegraphics[width=0.40\textwidth]{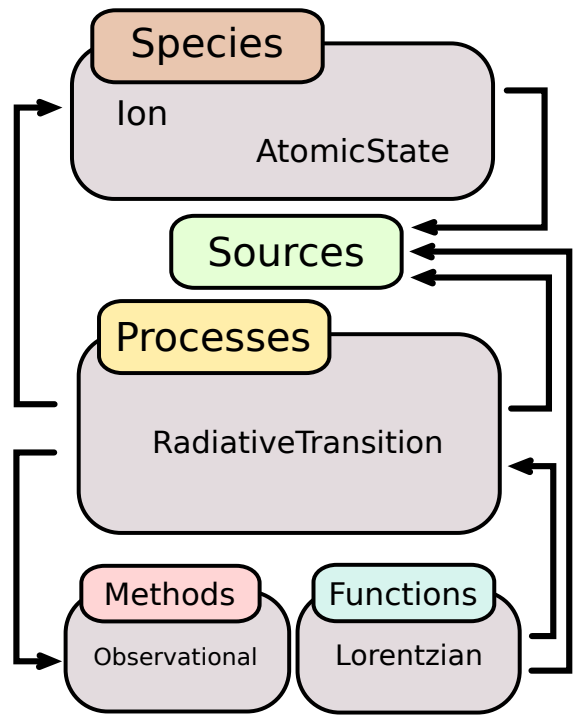}
  \caption{Simplified summary of XSAMS top level structure (colored
  boxes) with a small selection of the possible properties representing an
  atomic radiative transition (gray boxes). The depicted relationships (arrows) 
  hide a lot of complexity; for example the radiative transition references
  the species but also individual upper/lower states. Almost all properties can also
  reference its own sources.} \label{fig:xsamstoplevels}
\end{figure}

Data returned by a node must conform to the {XML Schema for Atoms, Molecules and Solids} (XSAMS).
XSAMS originated at the International Atomic Energy Agency (IAEA) and is now
maintained and further developed by VAMDC in collaboration with NIST. XSAMS
defines a strict layout for where and how data is represented in the XML document.
This includes for example how to describe atomic and molecular structure,
processes like collisions or radiative transitions, as well as supplemental
information such as environmental properties. In addition, associated
references are included and cross-referenced within the XML structure. The top
levels of the XSAMS structure are shown in a simplified form in Figure
\ref{fig:xsamstoplevels}.

XSAMS documents are validated against the schema. This ensures that the document
is self-contained with consistent referencing (no dangling cross-references) and that
no essential information is missing. The schema makes heavy use of its
own types and classes to make sure that properties like
data accuracies and evaluations are always represented in the same
way. 

VAMDC publishes its standards in an online \textbf{dictionary}\footnote{\url{http://dictionary.vamdc.eu/}}.
The \textbf{returnables} section of the standard defines exactly how a node returning XSAMS
should name a specific data property as well as what unit and data
type should be used. For example, the life time of an atomic state
must be identified as \emph{AtomStateLifeTime} and be a float given in
seconds.

\subsection{The query language: VSS2}
\label{sec:query_language_vss2}

The query language that data nodes understand is called {VAMDC SQL-subset 2}
(VSS2).  It consists of severely restricted SQL-like \texttt{SELECT}
statements.  VSS2 supports querying with binary operators \texttt{<=,>,\ldots},
\texttt{LIKE} and \texttt{IN} as well as logical relations with \texttt{AND} or
\texttt{OR}. Other standard SQL features, like  \texttt{BETWEEN}, are not included 
in the SQL-subset.  Instead of database column names, SELECT statements
use keywords from the VAMDC standard dictionary in a similar way as described for XSAMS in 
Section \ref{sec:output_language_xsams}.

Below is a full example of a query for atomic lines in FeIII:

\begin{verbatim}
    SELECT * WHERE RadTransWavelength < 858.4
       AND RadTransWavelength > 858.1
       AND AtomSymbol = 'Fe'
       AND AtomIonCharge = 2
\end{verbatim}

Note that there is no SQL FROM-clause in the above query. This is because the internal
table structure is hidden from users. The query is written as if the data existed in one
large in-line table. The keys used in the WHERE-clause in place of column names are
called \textbf{restrictables}. Restrictables are derived from the XSAMS itself and thus
generally correspond to a certain location in the data hierarchy. While the definition
of a Restrictable is fixed, it is up to each node to decide how to go about supplying
it. For example, a given node may need to do convert units or perform an internal query 
against more than one database column or table. 

It is important to note that, since XSAMS needs to be a self-contained
document, one cannot expect to receive only the isolated data type specified in
the SELECT-clause. That is, if the radiative transition we seek references a
certain state which in turn refers to a certain ion, \emph{both} the state and the ion
must be included in the returned document for it to be valid XSAMS. The VAMDC
dictionary however specifies a select number of \textbf{requestables}. One or
several such requestables can replace the \texttt{*} wildcard in the query to
request only a certain type of data. For instance, using the "Species"
requestable in \texttt{SELECT Species WHERE...} will only return the atoms and
molecules part of XSAMS. Leaving out the \texttt{WHERE} altogether will return
all species that a node has data about.

The exact same query can be sent to any VAMDC data node. It is the node's task
to convert the standard keys into a query relevant to the node's internal database. Each
node also needs to specify and report just which keys are actually relevant for
querying its data. This allows clients (like the VAMDC portal) to send a query only 
to nodes that can answer it, instead of gathering uncessary empty results.

\subsection{The API: VAMDC-TAP}
\label{sec:api}

A data node communicates with the world via HTTP, using GET/POST/HEAD requests.  The definition of
this web-API is known as {VAMDC-TAP}, where TAP stands for Table Access Protocol. This, in turn, is
a subset of the International Virtual Observatory Alliance TAP standard
(\citet{2012ivoa.spec.0827D}). VAMDC-TAP is a RESTful API that operates on the basis of a single
request and reply.  No sessions are stored or maintained. There is no connection between subsequent
queries and there is no authentication.

At the specified URI endpoint data nodes accept requests (HTTP GET/POST)
containing a VSS2 query as described in Section \ref{sec:query_language_vss2}.  If the
request is instead made using HTTP HEAD, the node answers 
with statistics about the given query (in the form of HTTP headers). Such statistics include
the number of species and processes that match the query as well as the expected size of the selected dataset. This
allows gathering information without actually executing a potentially expensive query and
subsequent big data transfer. This mechanism is used by the VAMDC Portal to gather
preliminary statistics from all nodes. Nodes can choose to truncate the data if the
result would be too large, reporting the truncation as part of the header return.

This is an example of a URI that uses VAMDC-TAP to query a data node: 

\begin{verbatim}
http://<node-url>/tap/sync?
                LANG=VSS2&
                REQUEST=doQuery&
                FORMAT=XSAMS&
                QUERY=<VSS2-query>
\end{verbatim}

The URI specifies the language, return format and the type of request. The
QUERY parameter holds an SQL-like string as specified in Section
\ref{sec:query_language_vss2}.
This example accesses the \texttt{tap/sync} endpoint to initiate a synchronous
query to the node, meaning that the data is returned right away, in the
response to the same HTTP request. There is currently no asynchronous
alternative to avoid blocking the server in the case of very expensive queries. This 
is planned for the next iteration of VAMDC-TAP and node software releases.

The VAMDC-TAP definition includes the additional administrative URI endpoints
\texttt{/tap/availability} and \texttt{/tap/capabilities}. The former allows
the node to report planned node maintenance or downtimes.  The node's
capabilities, a short XML document, include information like the URL addresses
to various services, the version of the node software used as well as sample
queries for testing the node in question. The capabilities document also
specifies which \textbf{returnables} (see Section \ref{sec:output_language_xsams})
and \textbf{restrictables} (see Section \ref{sec:query_language_vss2}) this node supports.  
This API can be accessed by tools like the VAMDC portal as well as by
custom use-case-specific applications. It is also intentionally easy
to script. 

VAMDC offers {processor} services. A processor takes an URI with a query as input, like in 
the example above. It then fetches the data and processes it before presenting the
result to the user. A complication arises in the form of the {same-origin policy} 
that web-browsers implement for security reasons. This means that browsers
don't make cross-site requests unless the website or service explicitly annouces that it
is fine with accepting them. Since VAMDC is inherently distributed and the various parts of the
infrastructure are meant to interact with each other at the user's behest, data nodes thus need
to announce their willingness to take cross-site requests. This is done by answering HTTP
SERVICE requests with the appropriate HTTP headers according to the cross-origin resource
sharing (CORS) standard\footnote{https://www.w3.org/TR/cors}.

\section{Node software implementation}
\label{sec:node_software_implementation}

From the description of the VAMDC Standards above, it should be clear that
there are obvious advantages with data nodes sharing a common software. The
nodes need not only answer to the same API -- incoming queries are also
node-independent, as is the data output format that needs to be assembled for
the reply.  However, depending on the type of data they hold, nodes can be
quite different internally. This concerns mainly the table layout of the
underlying database and is reflected in the list of restrictables and
returnables they support.

There are currently two implementations of the VAMDC standards for data nodes.
One is built on the Python programming language while the other is using Java.
All but two of the currently $\mathrm{\sim}$30 registered data nodes use the
Python node software which is the subject of this paper. The Python node
software can be obtained from its online git
repository\footnote{\url{https://github.com/VAMDC/NodeSoftware}}. It is
released under the GPL version 3 open-source license.

Python is an open-source high-level language with a mature ecosystem of utility libraries.
It is widely used within e.g.~the astronomical science community and tends to be fast to
learn for users with any previous programming experience. This has in our experience
lowered the barrier of entry for data providers wanting to start up their own node.

\subsection{Django and Data Model Abstraction}

The Python implementation of a VAMDC Node relies on the Django web
framework\footnote{\url{https://www.djangoproject.com}}. Django is widely used
as the underpinnings of professional large-scale websites, is actively
developed and has an active community.  It features a powerful and flexible ORM
(Object-Relational Mapping) which allows us to define a
database schema and formulate queries to it using Python rather than SQL. This
abstraction allows Django to support a wide range of different relational
database engines. In effect, data providers already storing their data in an
SQL database will likely find that Django can make use of it. 

The database engine is not formally part of the node software, but MySQL with
its MyISAM storage engine is used at many nodes. This also is the setup that has
been recommended to new nodes that do not already have their data in a
relational database. The recommendation still holds even though MySQL now uses InnoDB as
default storage engine, because in our use-case with de-facto read-only access
to the database, MyISAM has performance advantages once it has been set up.

Django's abstraction layer between the database and the node software API implementation
allows us to write code that is agnostic about a single node's intricacies. This allows
a large portion of the code to be reused by all data nodes. When a
node has special needs that makes the common implementation
insufficient, we add hooks where node-specific code can be plugged in.

The node software only uses a sub-set of Django's capabilities. Still, it
allows us to keep the code volume of the node software small: The core
functionality is only some 1000 lines, plus another 2000 for the
XML-generator. This has made maintenance and upgrades over the last seven years
relatively easy, even when seen over many major Django versions. Similarly, the
effort of porting the node software from Python 2.x to Python 3.x was measured
in hours rather than days; both are now supported with the same code-base to
ensure a smooth upgrade path for nodes.

The choice of Django has turned out to have a beneficial impact on several data
nodes, beyond their involvement with VAMDC. Django's relative ease of developing new custom
interfaces for data collections has allowed nodes to significantly improve their
own web presence, often heavily tailored to their pre-existing user communities. Examples of 
this are seen with for example CDMS\footnote{\url{http://cdms.berkeley.edu}}
(\citet{2005JMoSt.742..215M}) and HITRAN\footnote{\url{https://www.cfa.harvard.edu/hitran}}
(\citet{2013JQSRT.130....4R}).

\subsection{Publishing Tools}
\label{sec:publishing_tools}

Some atomic and molecular data are produced through ab initio calculations of atomic and
molecular energy structure, transition probabilities and collisional cross-sections etc.
Other data come from experimental measurements of wavelengths, line strengths and
lifetimes. Both approaches have advantages and problems. Occasionally, data producers 
merge the two e.g.~by combining experimental lifetimes with theoretical
branching ratios. In all cases, the new data comes in form of tables, often in peculiar
units and with complex assessment of the quality.

A stand-alone part of the node software, dubbed the \textbf{publishing tools},
parses such data and populates a new relational database for use by the VAMDC
node. Figure \ref{fig:vamdcstructure} shows the top-level structure of the
XSAMS output language the node is expected to return.  For query efficiency the
node should ideally try to organize its schema in a similar way.  For example,
the {Atom}, {Molecule} and {Transition} data should aim to be
separate database tables.

In the simplest case, the node data provider already has an SQL database with a
useful schema. Django's native \texttt{inspectdb} mechanism can use the existing
database for creating the Django data model. These auto-generated Python
classes usually only need small modifications before the node software can
access the database properly. In particular, the foreign key relations between
tables have to be transferred to the Django model.

However, in many cases the original data is either stored using some non-SQL
solution or using a schema less optimized for VAMDC
queries. This is where the publishing tools come into play: they can perform 
various operations on the data before inserting it into the node SQL database.
The goal is to do as much work as possible at database creation, when
execution time does not matter, instead of repeating work every time an
incoming query is processed. Possible operations include, but are not limited
to, proper parsing of near-arbitrary table formats, interpretation of header
information, tagging of missing data, unit conversions and pre-creation of derived
quantities that are useful in the XSAMS output.

The first step for doing this data import is to define the new (empty) database
schema. The new schema is formulated using
Django's model classes. Each such class represents a table in the new
database, including the relationships between them. Every instance of these
classes is one record in the table. This model is then used to create the actual
tables in the database, again granting nodes the flexibility to choose any of the
database engines that Django supports.

The mapping of data from the old data store to the new database can be the most
work-intensive part of setting up a VAMDC data node, depending on the original
form of the data. This is done by writing a \textbf{mapping file} that contains a
Python dictionary in a form understood by the conversion tool. It must be
created uniquely for each node. The components of the dictionary describe how
to read and parse the records of each input file into one or more output files.
Multiple inputs can for example be combined into a single output or
vice versa.  It is often not possible for the parser to work in parallel with
all input files; for example, inter-table relations may not be possible to
construct until all relevant files have been read. This requires multiple
passes, something which is supported by the mapping mechanism. Since the mapping file can
contain Python code, this process can be further customized with helper
functions to account for any peculiarities in the input data. The publishing
tool includes a set of functions for processing and handling
common data formats.

Running the mapping file through the execution script produces
a series of text files that exactly match the database schema -- one file per
database table with matching columns. These files can be efficiently read using
the database's own import functions. The syntax of this import is the only part
of the node creation that depends on the particular choice of SQL database.
For example, a user of MySQL would open the database client and execute:

\begin{lstlisting}[language=sql]
    mysql> LOAD DATA INFILE
        filename INTO TABLE table;
\end{lstlisting}

Once the mapping file exists, it is easy to re-run and tweak the
process, or to add new data to the node at a later time. The publishing tools
are not part of the day-to-day operations of the VAMDC node.

With the data loaded in the new database, one can set up addtional
database indices or perform other optimizations. This bit highly depends on the
individual node's database schema, size and engine.

Updates to the database schema often coincide with an update of the data
itself, meaning that it can be convenient to simply use the publishing tools
to recreate the entire database from an updated mapping file and input data files. 
If this is not desired or possible, Django comes with a very flexible system for database
migrations, using small Python files that describe the changes to the database
over time. These are executed in sequence by Django and, if written correctly,
can even be used to revert changes and roll back to a previous version of
the database in case of errors.

\subsection{Receiving and executing queries}
\label{sec:queries}

One of the two main tasks of the node software is to receive queries of the
form described in Section~\ref{sec:query_language_vss2} above, and execute them on
the node's database.

In order to do this, the node software needs to be told which VAMDC
{Restrictables} correspond to what column in which table of the database.  We
implement this in the most straight-forward way, a Python dictionay where the
key is the Restrictable and the value is the name of the Django model and the
relevant data field, concatenated by an underscore. This is the Django-internal way of
referring to model fields and we refer readers to the excellent Django
documentation for more information. An example of this restrictables dictionary
is shown in Section~\ref{sec:vald}. The point is that this dictionary
connects the (node-independent) Restrictable keywords to the (node-specific)
names of the right table and column in the database.

An incoming VSS2 query is validated as a first step. We use the
third-party package \texttt{pyparsing}\footnote{This is the node software's only
additional dependency apart from Django and the Python driver for the
database engine in question.} to formulate an SQL-dialect strictly limited to the
supported SQL subset. Rejecting any SQL statements except SELECT right
away also offers protection from a large set of potential malicious queries.

The parsed and validated query is then handed to what we call a node's
\textbf{query function}. This is the only piece of Python logic that is custom
to each node. The query function's goal is to turn the input query into
\texttt{QuerySets}. A queryset are Django's internal objects representing the
\emph{result} of an SQL query. A queryset is \emph{lazily evaluated}, meaning
that the query is only executed on the database once the result values are
accessed. This allows the query function to quickly set up the result and pass
it on to the XSAMS generator which is described in
Section~\ref{sec:xsams_generator} below. Only then, as we describe below, will the
database deliver data that get wrapped in XML and streamed directly to the
user that initiated the query. 

Nodes have, by design, near-total freedom on how to write the query function. 
Covering all of the complexity and subtleties of \AM\ data
collections would not be possible. Thus the only
constraint the system impose is that a node's custom query function returns the appropriate
QuerySets. Documentation and examples are provided to help data providers with
this step and most nodes use the method described in the following.

The node software offers a utility to convert a (parsed and validated) query
into a Django \texttt{Q-object}. Most nodes use this first thing in their query
function. A Q-object is Django's way of representing a \emph{database-agnostic}
query. It is suitable for direct interaction with Django's ORM. 

The next step is to apply the Q-object to the data model. This is where nodes
differ largely. For example a node that contains atomic transition line lists
is likely to first restrict on the transition model. It would then use the ORM to
figure out the corresponding species and states along with the references that need
to be attached. Other nodes might restrict the species first or use a totally
different strategy.

As mentioned above, the query function returns QuerySets that can
be used to populate the output XML document. Each QuerySet needs to correspond
to a top-level organizational group of XSAMS, like Atoms, Molecules,
Transitions, Collisions, Sources and so on.  The return value of the query
function is therefore a Python dictionary where the keys identify the part of
XSAMS to be used, and the values are the (yet to be evaluated) QuerySets.
The names of the keys to be used are listed in the node software documentation.

The total overhead for receiving a query, parsing it and setting up the result,
i.e.~running a node's query function, is typically below 0.5 sec which is
negligible compared to the total response time, especially for large queries.
This means that nodes typically are responsive and start returning data soon
after the query is initiated.

\subsection{Generating XSAMS}
\label{sec:xsams_generator}

The second main task of the node software is to take the data that comes from
the database and put it into an XSAMS document
(cf.~Section~\ref{sec:output_language_xsams}). In the previous section we described
how nodes return QuerySets from their query function. These are passed directly
into the node software's \textbf{XSAMS generator}.

This generator is named such not only because it assembles the XSAMS data
output, but also because it is a proper Python \texttt{generator} in the sense
that it uses \texttt{yield} statements. Generators in Python are structures that
can be looped over without being fully in memory when the loop starts.
Expressions get evaluated only for the current loop element. This nicely
matches the input QuerySets; only what is currently needed for the next piece
of output gets fetched from the database and put into XML. Especially for large
data set this is not only much more memory efficient, it also means less of a
delay before the first piece of data is ready to return.

The main task of the XSAMS generator is to go through the differents parts
of XSAMS in turn, passing each loop through the QuerySet it
received. Each iteration of the QuerySet produces an instance of the
Django data model (a row in the database) and the generator accesses its 
fields' values to build up the
corresponding piece of XSAMS. We do not use an XML library or a document
object model for assembling the XML. For maximum control we instead use plain Python string
concatenations, placeholders and \texttt{''.join()} on lists of strings.

In order to know which attribute of the current model instance needs to be
mapped to which element or attribute of XSAMS, the generator needs the
Returnables dictionary. This is analogous to the Restrictables
dictionary from the previous section, but in the opposite direction. Again the
dictionary keys are the (node-independent) keywords; these are hardcoded in the
XSAMS generator to the proper XML-location. Since the list of Returnable keys
is derived from XSAMS, there is a one-to-one correspondence.

The values of the dictionary tell the generator how to access the (current
iteration) model instance to retrieve the wanted information. We use an
internal function called \texttt{GetValue()} for this, allowing nodes to
fill the Returnables dictionary values in several ways. Firstly, they can
contain a static string, useful when a value is the same for all data and not
even stored in the database (like units). Secondly, it can be
the name of a model field in which case its value gets retrieved. Thirdly, it can
be the name of a model field in a \emph{different} model, connected by a
foreign key relation; as common with Django, table traversal in models is
possible by simply concatenating fields and foreign keys with periods.

The fourth possible option for filling the dictionary value is the name of a 
custom method on the data model. \texttt{GetValue()} calls this
method and inserts whatever it returns into the current place in XSAMS. This
allows for maximal flexibility, but for performance reasons nodes are
encouraged to not use this if they can pre-calculate data instead.
Note that while Python's \texttt{eval()} is the easiest
way to implement \texttt{GetValue()}, we went to some length to avoid it
because this function gets called very often and needs to be speedy.

While the XSAMS generator is built to cover as many nodes' requirements as 
possible, it is also possible to replace the common routines by
custom versions at several levels of granularity. Some nodes attach a custom
method to a data model that returns large pre-assembled chunks of XML, others
choose to overwrite the top-level function of the generator altogether and
re-use only parts of it. In practical use nodes have often found that it has
been easier to create a custom routine rather than adapting the standard one
to also cover a highly specialized case. By deviating from the common code base only when
necessary, nodes keep the volume of custom code small.

Apart from Returnables and Restrictables, there is a third group of VAMDC
standard keys, the Requestables (see Section~\ref{sec:query_language_vss2}).  When
one or several of these are given in the query, instead of \texttt{SELECT *},
this means that only certain parts of XSAMS are asked for. The generator in
this case simply skips the execution of the other parts.

\subsection{VAMDC-TAP implementation}
\label{sec:vamdc_tap_generator}

The previous sections describe how queries are parsed and how the reply
content is assembled. What remains is the implementation of the VAMDC-TAP
API. Django offers a straight-forward way to connect Python code
to URL endpoints. Since the API is simple, a single \texttt{view} function per
endpoint is enough. The \texttt{capabilities} and \texttt{availability} (see
Section \ref{sec:api}) endpoints are implemented by rendering simple templates
with a few configuration variables, using Django's own template mechanism. For
example, the keys from the \texttt{Returnables} and \texttt{Restrictables} 
dictionaries go into
the \texttt{capabilities} to tell other parts of the VAMDC infrastructure which
keys are understood by the data node in question.

The \texttt{sync} endpoint receives and answers queries. All this view needs to 
do is to call the query validation, the node's query function and
then pass its result to the XSAMS generator. Note that due to the \emph{lazy
evaluation} of both the query and the generator, the main queries have not yet
run at this point and no XML is yet held in memory. Only when the XSAMS
generator is passed to Django's \texttt{StreamingHTTPResponse} does the whole
machinery start churning and gradually producing the XSAMS document for the client.
XSAMS documents can be hundreds of MB in size, this method has proven superiour 
for our needs, compared to early attempts with Djangos templates or XML libraries.

The view-function behind \texttt{sync} also needs to distinguish between GET and HEAD
requests, the latter allowing for returning statistics. In addition, incoming OPTION 
requests are answered according to CORS to allow cross-site requests
(cf. Section~\ref{sec:api}). Thanks to Django's high-level tools for handling
requests, the whole API implementation is only about 150 lines of code and
therefore easy to understand and maintain.

\subsection{Deployment}
\label{sec:deployment}

A data node's performance depends to a large extent on the underlying database
itself. However, the assembly of large XML documents can also
become computationally expensive, not the least because the XSAMS generator is
generalized for use by all nodes. Instead of spending effort on optimization, it 
has proven beneficial for several nodes
to replace the standard Python interpreter by \texttt{PyPy}\footnote{\url{http://pypy.org}},
an alternative interpreter employing just-in-time
compilation for large gains in speed. Django and the node software run nicely
with PyPy, including the recent PyPy3.

The node software must be run by an application server able to
serve data over HTTP.  Since data providers
generally host their nodes themselves, the setup should be light weight and
easy to manage.  We recommend to use
\texttt{gunicorn}\footnote{\url{http://gunicorn.org}}, an application server
with minimal configuration. It is also written in Python and a common choice for
deploying Django apps. Alternatives include Apache \texttt{mod\_wsgi} and
\texttt{uWSGI}. Nodes frequently choose set-ups that deviate from the
recommendation, for example due to integration with other services on the same
machine or local expertise with certain tools.

Furthermore, it is good practice to put a proxy between the application and the
client. Its most important task is to offer \texttt{gzip} compression to
clients (XML is highly compressible). The proxy can also do load
balancing and provide HTTPS/TLS. The most common and recommended choice of
proxy is \texttt{Nginx}\footnote{\url{http://nginx.org/en}} but again, data
providers often choose whatever webserver they have running already.
\texttt{Caddy}\footnote{\url{https://caddyserver.com}} is a relatively new
webserver that can act as proxy, do compression and automatically handle free
certificates for HTTPS, with minimal configuration effort.

\section{A data node in practice: VALD}
\label{sec:vald}

The Vienna Atomic Line Database (VALD, \citet{vald2011}) stores atomic and molecular transition
data of astronomical interest\footnote{\url{http://vald.astro.uu.se}}.  This
section describes the process and experiences of using the node software to
adapt VALD into a VAMDC data node.

VALD originally uses a custom binary storage method. Since the
original VALD database will continue operation alongside its
VAMDC counterpart, this data was extracted using VALD's own tools into
a series of text files:

\begin{itemize}
    \item The species and their properties (charge, composition,
        ionization energy, isotopic fractions, InCHi keys designation etc.
        Each record of this file also contain a unique species number.
    \item A list of transitions and the corresponding energy levels,
        sorted by wavelength. This file has two lines per transition containing
        transition probability, line broadenings and the upper and lower level 
        description (energy, Land\'{e} factor, coupling model, term designations
        etc).
    \item A list of the quantum numbers for each
        level in the same order as the previous file (J, L, S,
        parity etc).
    \item A list of the custom reference-ID strings used in VALD, tied
        to a number identifier.
    \item A list of all data collections included in VALD 
        identified by an id string. Often such compilation has an associated
        critical evaluation paper that contains valuable quality analysis work.
    \item A BibTeX file linking VALD's reference-ID
        strings to actual Bibtex entries.
\end{itemize}

Note that no less than three different files are involved in tracking
references making the import process quite involved.

The next step was to prepare the VAMDC database. As discussed in
Section \ref{sec:publishing_tools}, this is done using Django \textbf{models},
each representing a database table in the database. In the VALD
case, the most relevant models are:

\begin{itemize}
    \item \texttt{Species}, containing all species data like mass number,
        ionization energy and identifers. This model also holds a
        many-to-many relationship to other species if this is a part of a
        molecule.
    \item \texttt{State} describes all atomic or molecular states and references the Species involved,
        the energy, Lande factor and quantum mechanical properties. Also the literature sources
        for the main properties are referenced.
    \item \texttt{Transition} contains the wavelength for each Transition alongside references to the
        States and Species involved. It also refers to all sources used for the data and which line list
        the transition data comes from.
    \item \texttt{Linelist} represents the line lists and sources used for every Species. It also
        designates if the data was obtained through empirical observations, calculations etc.
    \item \texttt{Reference} stores the actual bibtex references for all data in the database.
\end{itemize}

This data model was then used by Django's \texttt{migrate}-mechanism to create
the tables in an empty database.  With the database schema in place we now used
the publishing tools' mapping system to map the raw data files of VALD to the
VAMDC schema. Due to the layout of the raw data files the mapper was set to
read several files simultaneously and collate the lines (or groups of lines)
from each file to build an output record. The first pass created the records
and the second established the relationships between the tables. For
efficiency, a separate run created the many-to-many relationships as extra
intermediary data files. 

An important re-organization of data was carried out at this stage. The
original data inlines upper- and lower states as part of the transition record.
The XSAMS data model instead 
separates states and transitions and lets the latter refer to the former. We
therefore needed to de-duplicate states and create several output files from
the same input file. The relation between the states and transitions also had
to be preserved by generating unique keys tying them together.

Since VALD is a large database with 250 million lines, this conversion process
is computationally quite heavy. Originally the idea was to have the
node software mapper create Django database instances directly. This turned out to be both
slow and memory consuming (since a full Python object must be created for
each record). Creating intermediary text files that match the database schema
(cf.~Section~\ref{sec:publishing_tools}) turned out to be considerably faster and
is now a part of the standard workflow. Furthermore we found that the import
process benefited greatly from the use of the PyPy just-in-time compiling
Python interpreter; we saw a speedup of up to a factor five. Consequently we
kept using PyPy also for the node's normal operation and recommend PyPy in the
node software manual.

At this point the mapping tool had created a series of large text files, one for each
database table. These files were read directly into the MySQL database using MySQL's own
load mechanism. This is fast as long as the database indices are not in place yet. Django can automatically create indices for certain fields but we chose to defer this until after loading the data. Then we also created additional database indices to speed up the most common types of queries.

With the database ready, we proceed with preparing the dictionaries to
enable communication with VAMDC. The restrictables dictionary looks like this:

\begin{verbatim}
    "AtomSymbol": "species__name",
    "AtomMassNumber": "species__massno",
    ...
    "Lower.StateEnergy": "lostate__energy"
    "Upper.StateEnergy": "upstate__energy"
    ...
\end{verbatim}

This allows the node software to convert incoming VSS2 request into a Django
Q-object.  On the left hand side are the standard VAMDC restrictable keys. The
right hand side contains the corresponding model field name in Django format,
using double-underscores \verb|__| to denote membership. In the example above,
a request for \texttt{AtomSymbol} data leads to querying the \texttt{name} field
of the \texttt{species} table.

Next is the query function. As decscribed above, it receives the query and
starts by converting the SQL into a Django Q-object.  Because VALD is primarily
a transition line list, our query strategy is to start from the
\texttt{transitions} model and pass to it the Q-object, resulting in the (not
evaluated) QuerySet for all matching transitions.

Reconstructing the QuerySets for states and references containing only the
records needed for the selected transistions turned out to be slow for this
particular database. So in order to prematureley evaluate the transitions-QuerySet 
we opted for a customized generator instead; The VALD node overrides the main function of the
XML generator. Our custom version is a variant of the standard one. It uses all
the sub-functions creating the XML but has one important optimization: It starts 
outputting the transition data right away, \emph{before} figuring out which species, 
states and references will be needed to make the data complete. During the loop 
over transitions, we collect the ID-keys in Python \texttt{set}s. These sets are then
used in simple and fast database queries to retrieve the complementary data. For
queries close to the size limit this reduced the response time of the VALD node by 
up to a factor ten.

A drawback of this approach is that the VALD node no longer knows how much data
it will include in a given request and cannot include statistics in the GET HTTP headers.
To get the statistics headers from VALD an explicit HEAD request is needed. This
was deemed to be an acceptable compromise for the increase in speed.

The VALD \texttt{returnables} dictionary used by the XSAMS generator to populate the
XML looks like follows, making use of the full range of options described in
Section~\ref{sec:xsams_generator}:

\begin{verbatim}
   ...
   "AtomStateID": "AtomState.id",
   "AtomSymbol": "Atom.name",
   ...
   "RadtransSpeciesRef": RadTran.get_wave_refs()",
   "RadtransWavelengthUnit": "A",
   "RadtransUpperStateRef": "Radtran.upstate_id",
   "RadtransLowerStateRef": "Radtran.lostate_id",
   ...
\end{verbatim}

The VAMDC keys (on the lefthand side) tell the XSAMS generator where in XSAMS
to place a certain value. The dictioanry values are the constants, model fields or methods 
returning the content to insert.
Note that we make use of Django's features to access inter-table foreign keys through \texttt{\_id} and use a custom model method for the non-trivial way that VALD handles references.

After this, the VALD node was operational and could be registered with the VAMDC network.
An important aspect of the setup is that it's easily repeatable and scriptable: When a new
line list is added or a correction is done to VALD one can just re-run the process and get a
VAMDC version with the latest update.

\section{Concluding remarks}
\label{sec:concl}

We have in some detail described the VAMDC Standards that concern data nodes,
and the implementation of these protocols in practice. The software package
that has been the topic of this paper strikes an illustrative balance between
diverging requirements from different research groups and the goal to provide
common tools that make common tasks easy. It is also a case where a successful
web framework like Django becomes highly useful in science.

In summary, the node software carries out the tasks of
\begin{itemize}
\item Helping data providers importing existing data to a standard relational
database schema, starting from almost arbitrary custom storage
formats.
\item Making it easy to repeat the process when set up once. Such situations
occur when the data is updated/expanded or a new type of data is added to a
node.
\item Translating the VAMDC standard
query language (VSS2) to the internal names/fields/units conventions at a given
node.
\item Converting the data from the node's database to the VAMDC common output
format (XSAMS). 
\item Providing the API access needed to make a data node operate within 
the VAMDC infrastructure, including registry information.
\end{itemize}

Since its inception, the node software has been developed by a small team of
VAMDC consortium members who each are in charge of a data node.
This means that developers and users overlap significantly and that flexibility
and easy modification were always prioritized over user-friendliness. One can 
argue that "end users" wanting to use VAMDC to retrieve data never need to use or
understand the node software; it is used by data providers only. Nevertheless,
significant effort has been spent into documentation, both in text and screencast
video form, allowing new data nodes to be set up without significant input or help
from the node software developers.

Since its initial launch, there have been several node software releases,
marked with Git(Hub) tags. These follow the evolution of the VAMDC standards
and have also often offered gradual but significant improvements driven by
feedback from the data provider community or new requirements by an individual
data node. These custom changes are merged into the main version control branch
which is today considered a rolling beta release. The VAMDC standards
themselves are stable and currently only evolve in a backwards-compatible
manner. This means that unless a data node wants to make use of the latest
features, it needs not upgrade in order to stay interoperable with the
surrounding infrastructure. This fits well with the reality of data nodes; they
are often run with very little manpower for oversight, maintenance and
upgrades. To our knowledge this has not yet presented any problems, neither in
terms of interoperability nor in security.  Nevertheless we are currently preparing
a release based on the latest long-term-support version of Django (1.11, April
2017). After release we will encourage all data nodes to upgrade the node
software and the underlying stack to the latest version.


\section{Acknowledgements}
\label{sec:acknowledements}

The authors acknowledge the contributions, testing efforts and feedback by many
VAMDC collaborators that have benefited the node software tremendously. In
particular Guy Rixon, Nicolas Moreau, Christian Hill, Christan Endres, Michail
Doronin, Yaye-Awa Ba, and Johannes Postler.


\bibliographystyle{aa} 
\bibliography{citations.bib} 

\newpage
\onecolumn

\section*{Appendix: XSAMS example}

This is an annotated example of an XSAMS document (See Section
\ref{sec:output_language_xsams}) returned from a request of Ca lines
between 5000.0 and 5000.1\,\AA. The return data is from the VALD
database node. The output was shortened in places for brevity.

\lstset{language=XML}
\begin{lstlisting}
<?xml version="1.0" encoding="UTF-8"?>
<XSAMSData xmlns="http://vamdc.org/xml/xsams/1.0" 
    xmlns:cml="http://www.xml-cml.org/schema" 
    xmlns:xsi="http://www.w3.org/2001/XMLSchema-instance" 
    xsi:schemaLocation="http://vamdc.org/xml/xsams/1.0">
  <Processes>
    <Radiative>
      <RadiativeTransition id="Pvald-R571145" process="excitation">
\end{lstlisting}
This return concerns a radiative transition, which is identified using an ID
internally consistent within this XSAMS document. Other parts of the document
can then reference this as needed. 
\begin{lstlisting}
        <EnergyWavelength>
          <Wavelength>
            <Comments>Vacuum wavelength from state energies (RITZ)</Comments>
            <SourceRef>Bvald-K07</SourceRef>
            <Value units="A">5000.04125034</Value>
          </Wavelength>
        </EnergyWavelength>
\end{lstlisting}
The wavelength and its unit is given. It \emph{could} reference a
\texttt{Method} to show it is a calculated result but in VALD's case this
is instead noted as a comment and the reference to the source is
given (later in this document). 
\begin{lstlisting}
        <UpperStateRef>Svald-26696</UpperStateRef>
        <LowerStateRef>Svald-44769</LowerStateRef>
        <SpeciesRef>Xvald-191</SpeciesRef>
        <Probability>
          <Log10WeightedOscillatorStrength>
            <SourceRef>Bvald-K07</SourceRef>
            <Value units="unitless">-3.607</Value>
          </Log10WeightedOscillatorStrength>
        </Probability>
\end{lstlisting}
Similarly, the transition will reference other parts of the XML
document (the \texttt{AtomicState}) for the information about the
upper/lower states involved in this transition. The transition 
probability is given with values, units and reference. 
\begin{lstlisting}
        <ProcessClass />
        <Broadening name="natural" envRef="Evald-natural">
          <Comments>Natural Broadening</Comments>
          <SourceRef>Bvald-K07</SourceRef>
          <Lineshape name="lorentzian">
            <LineshapeParameter name="log(gamma)">
              <Value units="1/s">7.24</Value>
            </LineshapeParameter>
          </Lineshape>
        </Broadening>
\end{lstlisting}
Broadening is given with value and unit as well as with references to
sources and to the more detailed environment block later in the
document. This exemplifies natural broadening, other broadening types
(such as Van der Walls and Stark broadening) would follow (not 
shown here).
\begin{lstlisting}
      </RadiativeTransition>
    </Radiative>
  </Processes>
  <Species>
    <Atoms>
      <Atom>
        <ChemicalElement>
          <NuclearCharge>20</NuclearCharge>
          <ElementSymbol>Ca</ElementSymbol>
        </ChemicalElement>
        <Isotope>
          <IsotopeParameters>
            <MassNumber>40</MassNumber>
          </IsotopeParameters>
          <Ion speciesID="Xvald-191">
\end{lstlisting}
This describes the species (Isotope) that \texttt{RadiativeTransition} referenced above. 
\begin{lstlisting}
            <IonCharge>0</IonCharge>
            <AtomicState stateID="Svald-26696">
              <Description>3p6.3d.7f 3P*</Description>
              <AtomicNumericalData>
                <StateEnergy>
                  <SourceRef>Bvald-K07</SourceRef>
                  <Value units="1/cm">60690.2700</Value>
                </StateEnergy>
                <LandeFactor>
                  <SourceRef>Bvald-K07</SourceRef>
                  <Value units="unitless">1.41</Value>
                </LandeFactor>
              </AtomicNumericalData>
              <AtomicQuantumNumbers>
                <Parity>odd</Parity>
                <TotalAngularMomentum>1.0</TotalAngularMomentum>
              </AtomicQuantumNumbers>
              <AtomicComposition>
                <Component>
                  <Configuration>
                    <AtomicCore>
                      <Term />
                    </AtomicCore>
                  </Configuration>
                  <Term>
                    <LS>
                      <L>
                        <Value>1</Value>
                      </L>
                      <S>1.0</S>
                    </LS>
                  </Term>
                </Component>
              </AtomicComposition>
            </AtomicState>
\end{lstlisting}
The entirety of the Ion State \texttt{3p64d.7f 3P*} is described,
along with its quantummechanical data. The measurements of its 
energy and land\'e factor both come with references to which 
sources they are coming from. Note that more States would follow
after this one.
\begin{lstlisting}
            <InChI>InChI=1S/Ca</InChI>
            <InChIKey>OYPRJOBELJOOCE-UHFFFAOYSA-N</InChIKey>
          </Ion>
        </Isotope>
      </Atom>
    </Atoms>
  </Species>
  <Sources>
    <Source sourceID="Bvald-2017-02-20-14-58-15">
      <Comments>
        This Source is a self-reference. It represents the database
        and the query that produced the xml document.  The sourceID
        contains a timestamp. The full URL is given in the tag
        UniformResourceIdentifier but you need to unescape ampersands
        and angle brackets to re-use it.
        Query was: select * where 
                    (RadTransWavelength 
                        &amp;gt;= 5000.0 AND RadTransWavelength 
                        &amp;lt;= 5000.1) AND ((AtomSymbol = 'Ca'))
      </Comments>
      <Year>2017</Year>
      <Category>database</Category>
      <UniformResourceIdentifier>
        http://vald.inasan.ru/vald-node/tap/sync?LANG=VSS2&amp;
            REQUEST=doQuery&amp;
            FORMAT=XSAMS&amp;QUERY=select+*+where+%28RadTransWavelength+
            %3E%3D+5000.0+AND+RadTransWavelength+%3C%3D+5000.1%29+
            AND+%28%28AtomSymbol+%3D+%27Ca%27%29%29
      </UniformResourceIdentifier>
      <ProductionDate>2017-02-20</ProductionDate>
      <Authors>
        <Author>
          <Name>N.N.</Name>
        </Author>
      </Authors>
    </Source>
\end{lstlisting}
The node software will always return a "self reference" as part of the XSAMS document.
This details which query produced this result and is intended
primarily for the end user to be able to reproduce their query. It
could also potentially be used for bug reporting.
\begin{lstlisting}
    <Source sourceID="Bvald-K07">
      <Authors>
        <Author><Name>R. L. Kurucz</Name> </Author>
      </Authors>
      <Title>Robert L. Kurucz on-line database of observed and 
        predicted atomic transitions</Title>
      <Category>private communication</Category>
      <Year>2007</Year>
      <SourceName>unknown</SourceName>
      <Volume></Volume>
      <PageBegin></PageBegin>
      <PageEnd></PageEnd>
      <UniformResourceIdentifier>http://kurucz.harvard.edu/atoms/1400/gfemq1400.pos <!--...--></DigitalObjectIdentifier>
      <BibTeX>@misc{K07, Author = {{Kurucz}, R.~L.}, Title = {Robert L. Kurucz on-line 
        database of observed and predicted atomic transitions}, year = 2007, 
        Bdsk-Url-1 = {http://kurucz.harvard.edu/atoms/1400/gfemq1400.pos, ...}}
      </BibTeX>
    </Source>
\end{lstlisting}
The source reference contains all necessary components for properly
crediting the data, including an optional BibTex field. Note that parts of this
entry has been cropped.
\begin{lstlisting}
  </Sources>
  <Methods>
    <Method methodID="Mvald-0">
      <Category>experiment</Category>
      <Description>VALD exp - transition between levels with
                experimentally known energies</Description>
    </Method>
\end{lstlisting}
This is an example of the node providing more information than
necessary -- this method is not referenced from anywhere in the XSAMS
document but is still included (other methods have been cut out to
shorten the listing).
\begin{lstlisting}
  </Methods>
  <Functions>
    <Function functionID="Fvald-stark">
      <Name>Stark Broadening</Name>
      <Expression computerLanguage="Fortran">
            gammawaal * (T / 10000.0) ** (1.0/6.0) * N</Expression>
      <Y name="gammaL" units="1/cm3/s" />
      <Arguments>
        <Argument name="T" units="K">
          <Description>The absolute temperature, in K</Description>
        </Argument>
        <Argument name="N" units="1/cm3">
          <Description>Number density of neutral perturbers</Description>
        </Argument>
      </Arguments>
      <Parameters>
        <Parameter name="gammawaal" units="1/cm3/s">
          <Description>Lorentzian FWHM of the line</Description>
        </Parameter>
      </Parameters>
      <Description>This function gives the temperature dependence of
                Stark broadening.</Description>
    </Function>
\end{lstlisting}
The function block can offer functional code in computer language or other
representation. Multiple other functions would follow this one.
\begin{lstlisting}
  <Environments>
    <Environment envID="Evald-stark">
      <Comments>A given gamma can be scaled with
                gamma = gamma_given * (T / T_ref)^1/6 
                        * number density of free electrons.</Comments>
      <Temperature>
        <Value units="K">1.0E4</Value>
      </Temperature>
      <TotalNumberDensity>
        <Comments>The broadening parameters are given in
Hz per number density (i.e. 1/cm3/s), so they can simply
be scaled with the number density. Note that unless otherwise noted,
log10(gamma) is given.</Comments>
        <Value units="1/cm3/s">1</Value>
      </TotalNumberDensity>
    </Environment>
\end{lstlisting}
Again, additional environment sections were cut for brevity.
\begin{lstlisting}
  </Environments>
</XSAMSData>

\end{lstlisting}

\end{document}